\documentclass[12pt,preprint]{aastex}

\DeclareMathAlphabet{\mathsc}{OT1}{cmr}{m}{sc}
\def\testbx{bx}%
\DeclareRobustCommand{\ion}[2]{%
\relax\ifmmode
\ifx\testbx\f@series
{\mathbf{#1\,\mathsc{#2}}}\else
{\mathrm{#1\,\mathsc{#2}}}\fi
\else\textup{#1\,{\mdseries\textsc{#2}}}%
\fi}

\begin{document} 
\title{The Lesser Role of Shear in Galactic Star Formation: Insight from the Galactic Ring Survey}

\author{Sami Dib\altaffilmark{1}, George Helou\altaffilmark{2}, Toby J. T. Moore\altaffilmark{3}, James S. Urquhart\altaffilmark{4}, Ali Dariush\altaffilmark{1}}

\altaffiltext {1} {Astrophysics Group, Blackett Laboratory, Imperial College London, London, SW7 2AZ, United Kingdom; s.dib@imperial.ac.uk}
\altaffiltext {2} {Infrared Processing and Analysis Center, California Institute of Technology, Pasadena, CA 91125, USA}
\altaffiltext {3} {Astrophysics Research Institute, Liverpool John Moores University, Twelve Quays House, Egerton Wharf, Birkenhead CH41 1LD, United Kingdom}
\altaffiltext {4} {Max-Planck Institut f\"{u}r Radioastronomie, Auf dem H\"{u}gel 69, 53121, Bonn, Germany}

\begin{abstract} 

We analyse the role played by shear in regulating star formation in the Galaxy on the scale of individual molecular clouds. The clouds are selected from the $^{13}$CO $J=1-0$ line of the Galactic Ring Survey. For each cloud, we estimate the shear parameter which describes the ability of density perturbations to grow within the cloud. We find that for almost all molecular clouds considered, there is no evidence that shear is playing a significant role in opposing the effects of self-gravity. We also find that the shear parameter of the clouds does not depend on their position in the Galaxy. Furthermore, we find no correlations between the shear parameter of the clouds with several indicators of their star formation activity. No significant correlation is found between the shear parameter and the star formation efficiency of the clouds which is measured using the ratio of the massive young stellar objects luminosities, measured in the Red MSX survey, to the cloud mass. There are also no significant correlations between the shear parameter and the fraction of their mass that is found in denser clumps which is a proxy for their clump formation efficiency, nor with their level of fragmentation expressed in the number of clumps per unit mass. Our results strongly suggest that shear is playing only a minor role in affecting the rates and efficiencies at which molecular clouds convert their gas into dense cores and thereafter into stars. 

\end{abstract} 

\keywords{galaxies: ISM, ISM: clouds, ISM: molecules, stars: formation}

\section{Introduction}\label{motiv}

The rates and efficiencies at which galaxies convert gas into stars determine their evolution and their observable properties. The diffuse phase of the interstellar medium (with number densities in the range $\sim 0.1-1$ cm$^{-3}$) in galaxies is subject to a variety of instabilities such as large scale gravitational instabilities, shear effects from differential galactic rotation, and expanding superbubbles created by supernova explosions (Elmegreen 1995; McKee \& Ostriker 2007 and references therein). Compressive motions associated with these instabilities cause a transition to the molecular phase (with number densities $\sim 100$ cm$^{-3}$) and to the formation of molecular clouds . More massive clouds can also form by the collision of smaller mass ones (e.g., Tan 2000). The survival of molecular clouds may well depend on their ability to become self-gravitating before being affected by shear. As stars form in the densest regions of molecular clouds (e.g., Blitz 1993; Andr\'{e} et al. 2009), it is therefore of prime importance to assess the relevance of the physical processes that affect the evolution of the clouds and the rates and efficiencies with which they convert gas into stars. Supersonic turbulence which is observed ubiquitously in molecular clouds (e.g., Heyer \& Brunt 2004; Schneider et al. 2011) produces local compressions, a fraction of which can be 'captured' by gravity and proceed to form stars (e.g., Klessen et al. 2000; Goodwin et al. 2004; Dib et al. 2007; Dib \& Kim 2007; Offner et al. 2008; Kritsuk et al. 2011; Dib et al.  2010a; Csengeri et al. 2011; Padoan \& Nordlund 2011). Magnetic fields also play an important role in determining the fraction of gravitationally bound gas in star forming clouds. Results from numerical simulations show that stronger magnetic fields lower the rate of dense core formation (e.g., V\'{a}zquez-Semadeni et al. 2005; Price \& Bate 2008; Dib et al. 2008;2010a; Heitsch et al. 2009; Li et al. 2010). The regulation of the star formation rates on galactic scales have been explored through scenarios in which stars form as the result of gravitational instabilities in the disk (Madore 1977; Slyz et al. 2005; Li et al. 2006; Dobbs 2008). The gravitational instability can be mediated by thermal instabilities (Wada et al. 2000; Dib 2005; Dib \& Burkert 2005; Dib et al. 2006; Khesali \& Bagherian 2007; Shadmehri et al. 2010; Kim et al. 2011) and influenced by turbulence (Romeo et al. 2010; Shadmehri \& Khajenabi 2012). Wong \& Blitz (2002) and Blitz \& Rosolowsky (2006) argued that the fraction of star forming gas in galaxies is related to the pressure of the interstellar medium. The role of stellar feedback in regulating the star formation rates (SFRs) and efficiencies (SFEs) in molecular clouds has been highlighted in a number of recent studies (e.g., Murray et al. 2010; Dib et al. 2009;2011; Dib 2011). In particular Dib et al. (2011) and Dib (2011) showed that the SFEs and SFRs depend critically on the strength of the metallicity dependent, radiation driven winds. Weaker winds, associated with lower metallicities, allow for longer episodes of star formation in the clumps/clouds and lead to higher SFEs. 

Another physical agent that has been suspected of participating in the regulation of the SFR is the level of shear in galaxies (Silk 1997), or shear induced cloud collisions (Tan 2000). the model of Tan (2000) predicts an enhanced/reduced SFR in region of high/low shear. On the observational side, Seigar (2005) used observations for 33 nearby galaxies and argued for the existence of a correlation between the shear rates of the galaxies and their ratio of far-infrared to $K_{s}$-band luminosity which is a proxy for the specific star formation rates (SSFR). However, he found a very weak, insignificant correlation between the shear rate and the surface density of the SFR. As pointed out by Seigar, the stronger correlation between the SSFR and shear may simply reflect the correlation between the size of the optical disk and the K-band luminosity rather than between the SFR and shear. Relatedly, a recent study by Watson et al. (2012) for a sample of 20 bulgeless galaxies found no correlation between the SFEs of the galaxies and their circular velocity. Hunter et al. (1998) assessed the competition between self-gravity and shear in a number of Irregular galaxies. They found rather poor correlations between the shear strength and the SFRs. Elson et al. (2012) applied the Hunter et al. (1998) analysis to the blue compact dwarf galaxies NGC 2915 and NGC 1705. They found that the extent of the regions in which shear is important in these galaxies matches approximately the size of their stellar disks. However, they do not report on the quantitative relationship between the shear strength and the SFR. On the other hand, Weidner et al. (2010) presented numerical simulations of star cluster formation for clumps of masses $10^{6}$ M$_{\odot}$, sizes of $50$ pc, and of varying rotational support. They found that higher initial shear levels expressed in the form of initially larger rotational energies lead to a reduction of the SFE and of the SFR. Hocuk \& Spaans (2011) modelled star formation in molecular clouds within an AGN. They varied the level of shear by varying the mass of the black hole while keeping the cloud at the same distance. Hocuk \& Spaans also report a reduction of the SFE and SFR as the shear induced by the black hole increases. 

While the results of Weidner et al. (2010) and Hocuk \& Spaans (2011) show a clear trend in which the SFR and SFE may decrease with increasing shear levels, their simulations did not include magnetic fields nor feedback, both of which act to reduce the SFR and SFE. If lower levels of shear would allow for the formation of stars at higher rates, feedback from the first generation of stars formed in clouds  will disperse the remaining gas in the cloud on shorter timescales and thus, reduce the SFE (as compared to a case with no feedback). The observational characterisation of the role of shear in star formation (Hunter et al. 1998; Seigar 2005; Koda et al. 2009; Elson et al. 2012) has primarily investigated the ability of galaxies to convert diffuse gas into molecular clouds. While there is some observational evidence that high shear levels may prevent the formation of star forming molecular clouds (e.g., Elson et al. 2012), the role of shear in determining the SFR and SFE within these clouds is poorly understood. So far, there has been no observational tests to determine whether shear plays a significant role in star formation on the scale of individual clouds.

In this work, we use data from the Galactic Ring Survey (GRS, Jackson et al. 2006) in order to calculate the shear parameter, $S_{g}$, which compares the level of shear to self-gravity on the scale of $\sim 730$ clouds in the second quadrant of the Galaxy. We construct the distribution of the $S_{g}$ parameters and inspect the dependence of $S_{g}$ on Galactocentric radius. We search for correlation between the measured $S_{g}$ values of the clouds and a proxy of their SFE using the observation of massive young stellar objects (MYSOs) in the Red MSX survey (Urquhart et al. 2011). We also search for correlations between $S_{g}$ and the level of fragmentation of the clouds as well as with the mass fraction of the clouds that is found in denser clumps and which is a proxy for their core formation efficiency. In \S.~\ref{data} we briefly describe the GRS and Red MSX surveys and discuss the cloud selection. We describe the method that is employed to quantify the effects of shear in \S.~\ref{quantifshear} and in \S.~\ref{results} inspect the correlations between the shear parameter and the star formation efficiency indicators. In \S.~\ref{conclusions}, we discuss our findings and conclude.

\section{Data: The Galactic Ring Survey and the Red MSX Source Survey}\label{data}

\begin{figure}
\plotone{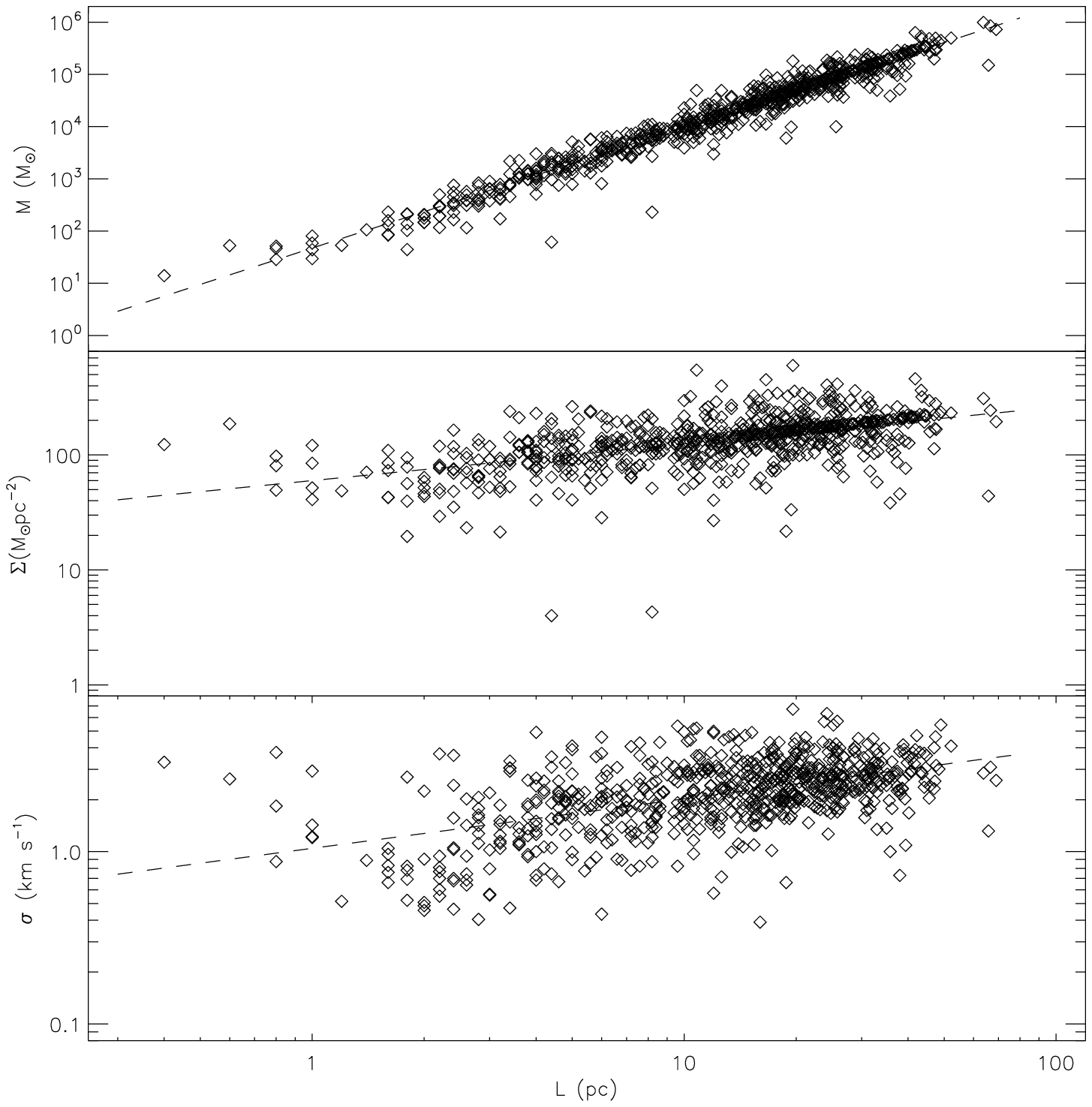}
\vspace{1cm}
\caption{Correlations between the masses of the clouds in the selected sample of GRS clouds and their sizes (top panel), their surface density and sizes (middle panel), and their internal velocity dispersions and sizes (lower panel). The lines in each panels are power law fits to the data points and their coefficients and exponents are quoted in \S.~\ref{data}.}
\label{fig1}
\end{figure}

The Boston University-Five College Radio Astronomy Observatory Galactic Ring Survey (GRS) is a $^{13}$CO $J= 1-0$ emission line survey which covers the Galactic longitudes $18^{\circ} < l < 55.7^{\circ}$ and Galactic latitudes $|b| \leq 1^{\circ}$ with a spectral resolution of $0.21$ km s$^{-1}$ and a spatial resolution of $46\arcsec$ (Simon et al. 2001; Jackson et al. 2006).  The velocity range of the GRS clouds starts at -5 km/s and so excludes most clouds outside the solar circle. Rathborne et al. (2009) applied an adapted version of the CLUMPFIND algorithm (Williams et al. 1994) to the $^{13}$CO data cubes and identified 829 clouds and 6124 clumps within them. Roman-Duval et al. (2009) measured the kinematical distances to $752$ clouds from the Rathborne et al. (2009) sample assuming the Clemens (1985) rotation curve for the Galaxy with the parameters ($R_{0},V_{0}$)=($8.5$ kpc, $220$ km s$^{-1}$), where $R_{0}$ is the Galactocentric distance of the Sun, and $V_{0}$ is the rotation velocity of the gas at the position of the Sun. The masses of 749 of these clouds, their surfaces densities, and their velocity dispersions have been derived by Roman-Duval et al. (2010). The clouds in the GRS are affected by the Malmquist bias (Roman-Duval et al. 2010) and the masses of the nearby, low mass, molecular clouds are highly uncertain. We therefore select clouds whose mass is $> 10$ M$_{\odot}$. We crossed matched the data of Rathborne et al. (2009), Roman-Duval et al. (2009), and Roman-Duval et al. (2010) and after eliminating a few clouds whose kinematical distances are degenerate, we obtained a sample which contains the masses of 727 clouds $M$, their surface densities $\Sigma$, velocity dispersions $\sigma$, sizes $L$, number of clumps $N_{cl}$, and kinematical distances $D$. The size of a cloud in the GRS is defined as being the diameter of a circular surface whose area is equal to the surface area covered by the cloud, $s$, and is given by $L=\sqrt{4 s/ \pi}$. Their masses fall in the range [$14-9.92\times 10^{5}$] M$_{\odot}$, their surface densities between [$4-601$] M$_{\odot}$ pc$^{-2}$, their velocity dispersions between [$0.38-6.70$] km s$^{-1}$, and their sizes between [$0.4-69$] pc, with median values of $2.54\times 10^{4}$ M$_{\odot}$, $143.9$ M$_\odot$ pc$^{-2}$, $2.27$ km s$^{-1}$, and $15$ pc, respectively. Fig.~\ref{fig1} displays the correlation between the masses of the clouds, $M$, their surface densities, $\Sigma$, and 3D velocity dispersions, $\sigma$, versus their sizes, $L$. Fits to the data displayed in Fig.~\ref{fig1} are made using the following scaling relations:

\begin{equation}
M ({\rm M_{\odot}})= C_{1} L^{\gamma} ({\rm pc}),
\label{eq1}
\end{equation}

\begin{equation}
\Sigma ({\rm M_{\odot} pc^{-2}})= C_{2} L^{\delta} ({\rm pc}),
\label{eq2} 
\end{equation}

and 

\begin{equation}
\sigma ({\rm km~s^{-1}})=C_{3} L^{\beta} ({\rm pc}).
\label{eq3}
\end{equation}

with $C_{1}=10^{1.67 \pm 0.021}, C_{2}=10^{1.77 \pm 0.020}, C_{3}=10^{0.018 \pm 0.019}, \gamma=2.31 \pm 0.018, \delta=0.32 \pm 0.018$, and $\beta=0.28 \pm 0.017$. A separate fit to the data points of gravitationally bound clouds (defined here as possessing a virial parameter $\alpha_{vir} <1$)\footnote{We employ the same definition of the virial parameter as in Roman-Duval et al. (2010), given by $\alpha_{vir}=(1.3 R \sigma/G M)$. Dib et al. (2007) pointed out that using $\alpha_{vir}$ overestimates the true gravitational boundedness of the clouds because of the neglect of other energy contributions (i.e., magnetic terms and surface energy terms) in the virial equation (see also Ballesteros-Paredes 2006 and Shetty et al. 2010 who pointed out how projection effects can influence virial parameters estimates). This implies that not all clouds that have $\alpha_{vir} < 1$ are truly gravitationally bound.}  yields the following values for the coefficients and exponents of the scaling relations: $C_{1}=10^{1.72 \pm 0.029}, C_{2}=10^{1.82 \pm 0.022}, C_{3}=10^{-0.241 \pm 0.024}, \gamma=2.28 \pm 0.024, \delta=0.295 \pm 0.023$, and $\beta=0.46 \pm 0.019$. The value of $\beta=0.29$ is smaller than the value of $0.5$ found by Solomon et al. (1987) for their selection of clouds in the same sector of the Galaxy in the Massachusets-Stony Brook Galactic plave CO survey. One possible origin for the discrepancy is that the GRS observations use the $^{13}$CO (1-0) line in contrast to the $^{12}$CO (1-0) line observations used by Solomon et al. (1987). The higher excitation density of the $^{13}$CO line allow for a finer separation of structures in the (l,b,v) space, and may be at the origin of the larger scatter observed in the bottom panel of Fig.~\ref{fig1} as compared to Fig. 1 in Solomon et al. (1987). We calculate the clouds Galactocentric radius, $R$, as being their Galactocentric distances projected onto the Galactic plane and which are given by: 

\begin{equation}
R^{2}=R_{0}^{2}-2 R_{0} D {\rm cos}(b) {\rm cos}(l)+ D^{2} {\rm cos}^{2}(b),
\label{eq4}
\end{equation}

\noindent where $R_{0}$ is the Galactocentric distance of the Sun which we take to be $8.5$ kpc, and $l$ and $b$ are the Galactic latitude and longitude of the clouds, respectively.

In the present work, we also make use of the catalogue of mid-infrared detected massive young stellar objects (MYSOs) from the Red MSX Source (RMS) survey of Urquhart et al. (2011). The RMS sample of MYSOs for the GRS clouds is complete to $L_{bol} > 10^{4} L_{\odot}$ out to a heliocentric distance of $\sim 14$ kpc and covers the Galactocentric radius range of $2.5$ to $8.5$ kpc. The sample consists of 176 RMS sources associated with 123 GRS clouds while the remaining GRS clouds have no matching RMS detection above $10^{4}$ L$_{\odot}$. The source luminosities were calculated by constructing the SEDs from various public data sources and fitting them with the YSO model fitter of Robitaille et al. (2006). The derived luminosities are effectively bolometric although dominated by the infrared data (more details can be found in Mottram et al. 2011).  

\section{Quantifying the Effect of Shear}\label{quantifshear}

Elmegreen (1993) and Hunter et al. (1998) argued that if condensations in the interstellar medium (ISM) accumulate mass because of streaming motions along magnetic field lines, then their growth rate is determined by the competition between their self-gravity and the local level of galactic shear. This competition will be more relevant than the one based on the competition between self-gravity, pressure, and the Coriolis force  (which is represented by the Toomre $\cal{Q}$ parameter, Toomre 1964) because magnetic fields can transfer the angular momentum away from the cloud (Elmegreen 1987). The growth of the density perturbations against shear is given by $(\pi~G~\Sigma) / \sigma$, where $\Sigma$ and $\sigma$ are the local gas surface density and velocity dispersion, respectively. The growth of the perturbation is most effective between $-1/A$ and $1/A$, where $A$ is the Oort constant given by: 

\begin{equation}
A= -0.5 R \frac{d\Omega} {dR}=0.5 \left(\frac{V}{R}-\frac{dV}{dR} \right),
\label{eq5}
\end{equation}

\noindent where $\Omega$ and $V$ are the angular and rotation velocities of the gas, respectively. The amplitude of the growth from an initial perturbation of the surface density $\delta\Sigma_{0}$ will be given by:

\begin{equation}
\delta\Sigma_{peak} \sim \delta\Sigma_{0} {\rm exp}\left(\frac{2 \pi G \Sigma}{\sigma A}\right).
\label{eq6}
\end{equation}

Hunter et al. (1998) argued that for a perturbation in the diffuse phase of the ISM to be significant and not to be erased by shear, it must grow by a large factor, $C$, which they chose to be $C=100$. A factor of $\sim 100$ corresponds to the density contrast between the diffuse phase of the ISM with densities $\sim 0.1-1$ cm$^{-3}$ and the molecular phase with densities $\gtrsim 100$ cm$^{-3}$. This leads to a critical surface density $\Sigma_{sh}$ given by:

\begin{equation}
\Sigma_{sh}=\frac{ \alpha_{A} A \sigma}{\pi G},
\label{eq7}
\end{equation}

\noindent where $\alpha_{A}={\rm ln}(C)/2$ (in Hunter et al. $\alpha_{A}={\rm ln}(100)/2 \sim 2.3$). One can then define a shear parameter for gravitational instability, $S_{g}$, which is given by:

\begin{equation}
S_{g}=\frac{\Sigma_{sh}}{\Sigma}=\frac{\alpha_{A} A \sigma}{\pi G \Sigma}.
\label{eq8}
\end{equation} 

\begin{figure}
\plotone{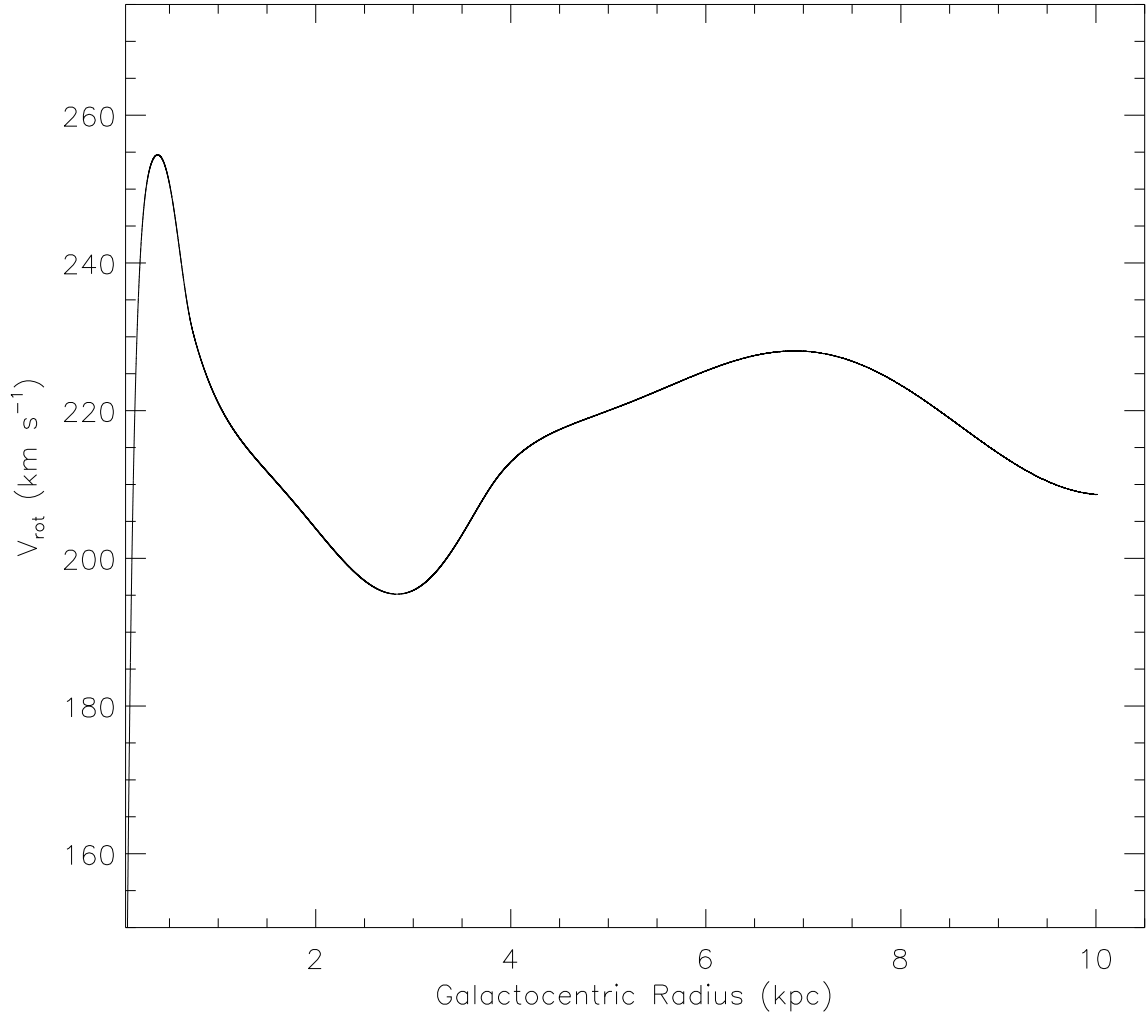}
\caption{Rotation curve of the Galaxy in the model by Clemens (1985) with the parameters of his model being set to [Galactocentric Radius of the Sun, Rotational Velocity at the position of the Sun]=[$R_{0}, V_{0}$]=[$8.5$ kpc, $220$ km s$^{-1}$].}
\label{fig2}
\end{figure}

\begin{figure}
\plotone{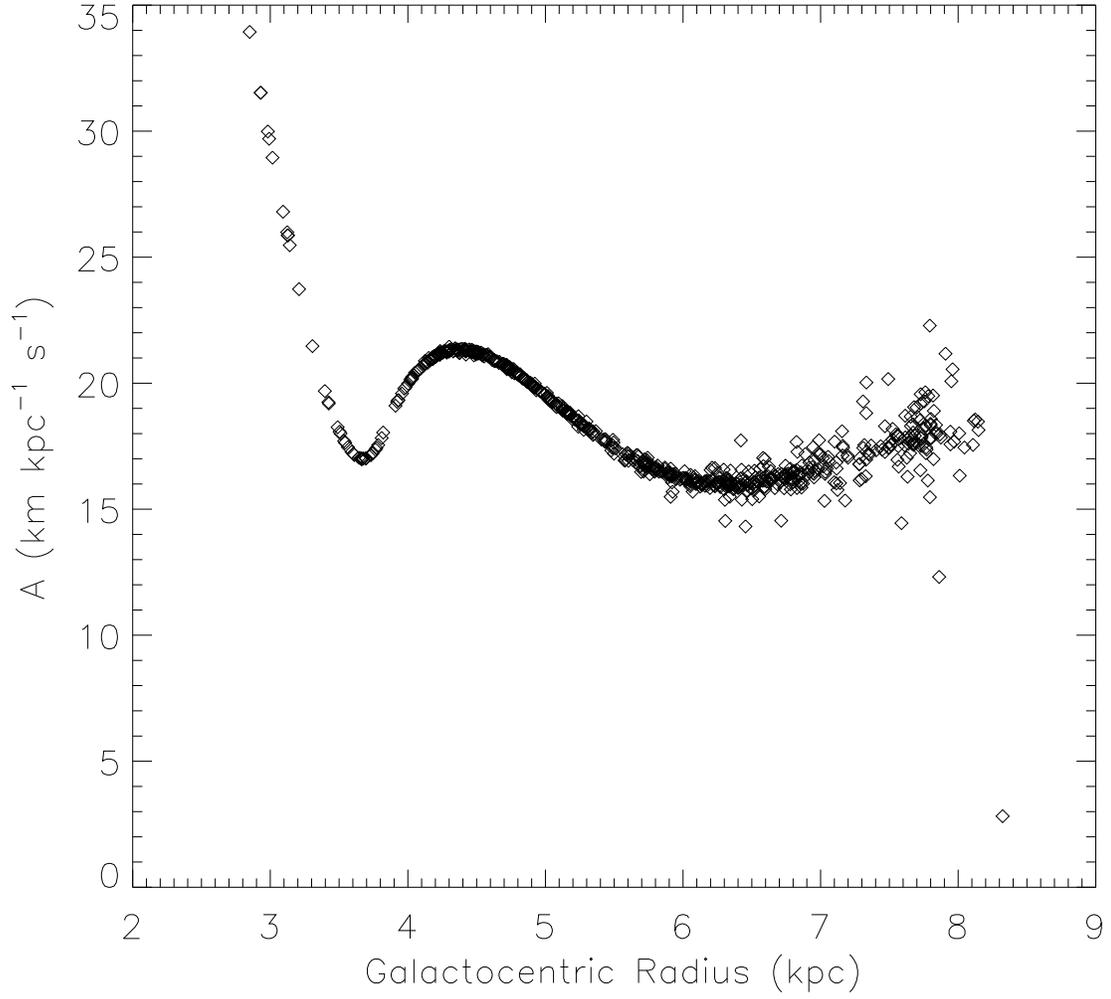}
\vspace{0cm}
\caption{Values of the Oort parameter calculated on the scale of individual GRS molecular clouds using the Clemens (1985) Galactic rotation curve.}
\label{fig3}
\end{figure}

\begin{figure}
\plotone{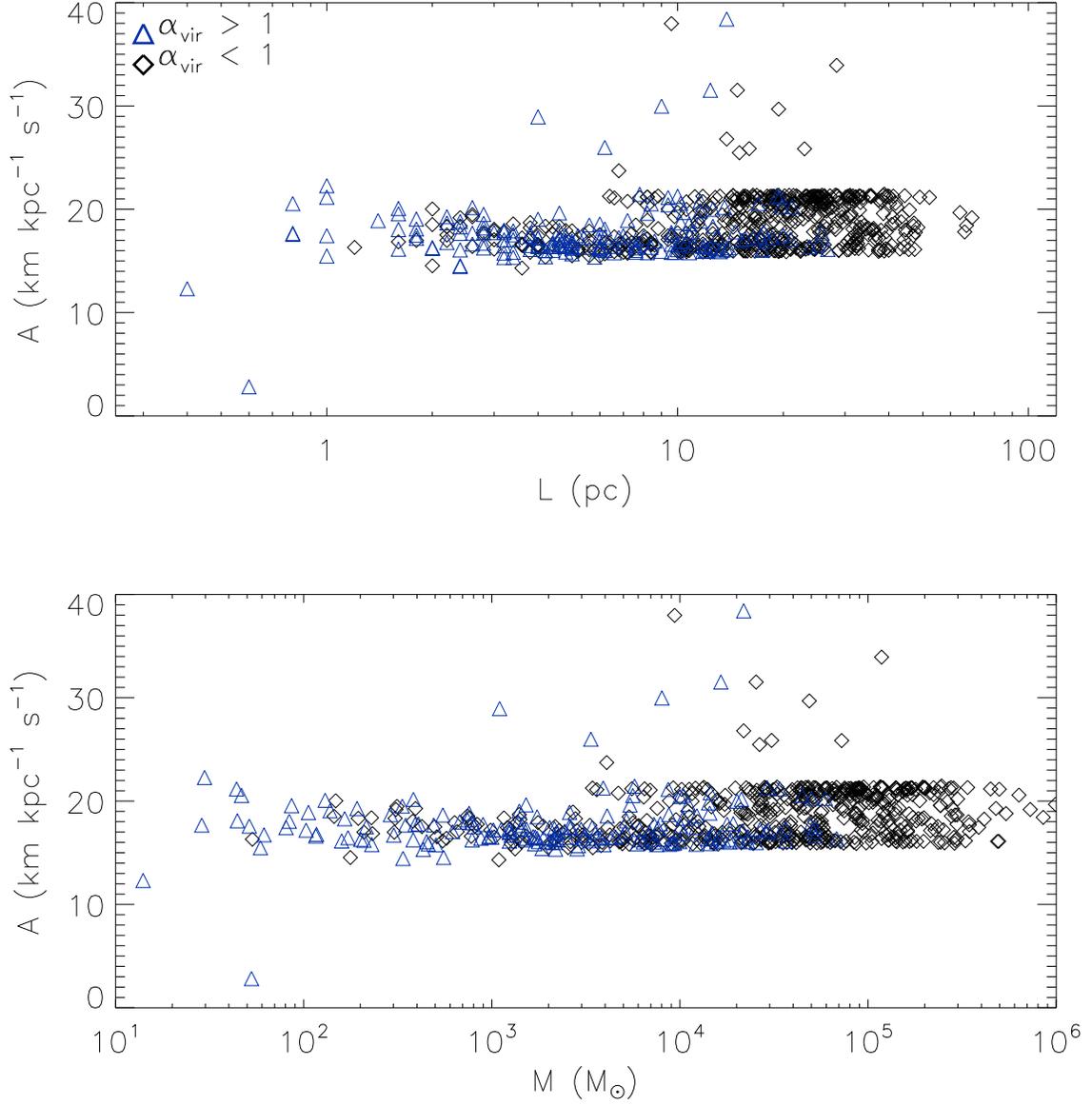}
\vspace{0cm}
\caption{Dependence of the Oort parameter values for the individual GRS clouds on their masses (bottom) and sizes (top). The black diamonds are for gravitationally bound clouds (defined as having $\alpha_{vir} \le 1$), and the blue triangles are for unbound clouds (with $\alpha_{vir} > 1$).}
\label{fig4}
\end{figure}

Shear will disrupt density perturbations when $S_{g}  > 1$ and would be ineffective in erasing them when $S_{g} < 1$. Measurements of  $S_{g}$ using \ion{H}{i} 21 cm line observations in a number of external galaxies on physical scales of $\sim 300-400$ pc by Hunter et al. (1998) and Elson et al. (2012) yield values that vary between a few times $0.1$ in the central regions and $\sim 10$ in the outer regions of their disks. We now apply the formalism described above to our sample of GRS clouds. For a cloud of size $L$, whose centre is located at the Galactocentric radius $R$, we calculate the corresponding quantity $A$ as being:

\begin{equation}
A=0.5 \left(\frac{V_{rot,c}}{R}-\frac{|V_{rot}(R+L/2)-V_{rot}(R-L/2)|}{L} \right),
\label{eq9}
\end{equation}

\noindent where $V_{rot,c}$ is the Galactic rotational velocity at $R$. The rotational velocities at the positions of the clouds centres are measured from the azimuthally averaged Galactic rotation curve of Clemens (1985), which is displayed in Fig.~\ref{fig2}. Ideally, the Galactic velocity gradient on a scale equal to the size of the cloud, ($|V_{rot}(R+L/2)-V_{rot}(R-L/2)| / L$), measured between the cloud's inner and outer Galactic edges could be estimated using the local \ion{H}{i} velocity curve. However, this information is not available for individual GRS clouds. It is also important to note that  this rotation curve represents a polynomial fit to discrete measurements of the rotational velocity. These measurements have an irregular spatial distribution and do not necessarily resolve physical scales that are of the order of the sizes of GMCs of a given size at any given Galactic longitude. An alternative would be to use velocity gradients derived in molecular lines for the individual GMCs. However, due to the high levels of compressibility of the molecular gas by the supersonic motions in the clouds, the global velocity gradients that are estimated using molecular lines observations tend to be altered from the local velocity gradient observed in the \ion{H}{i}  21 cm line. The former tend to be a few up to several times smaller than the latter (e.g., Rosolowsky 2007; Imara \& Blitz 2011; Imara et al. 2011). It is quite possible that measuring the $(dV/dR)$ term using the Clemens rotation curve may lead to an overestimate of the true velocity gradient for some of the clouds and an underestimate for others. This will generate a symmetric scatter in the derived values of $(dV/dR)$ for an ensemble of clouds of  similar properties located at a given Galactocentric radius. An additional reason for adopting the azimuthally averaged Galactic rotation curve is that the distances of the GRS clouds have been estimated by Roman-Duval et al. (2009) using the Clemens (2005) azimuthally averaged rotation curve of the Galaxy. Fig.~\ref{fig3} displays the values of $A$ for the selected sample of GRS clouds derived using Eq.~\ref{eq9}. Fig.~\ref{fig3} shows that for most clouds, the term $(dV/dR)$ is small and the value of $A$ is dominated by the term $(V/R)$. Fig.~\ref{fig4} displays the dependence of $A$ as a function of the cloud masses (bottom) and sizes (top). 

\begin{figure}
\plotone{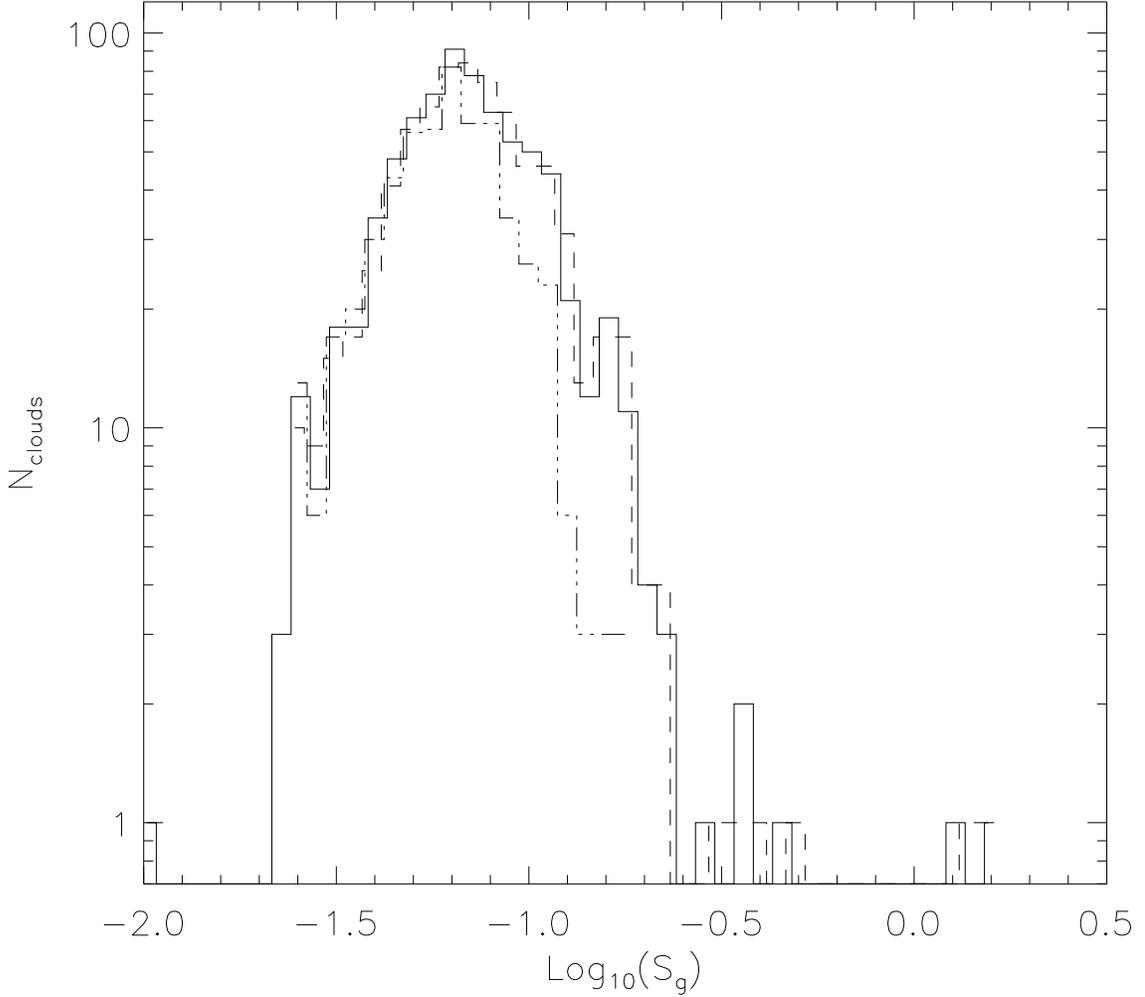}
\vspace{0cm}
\caption{Distribution of the shear parameter $S_{g}$ for the selected sample of molecular clouds in the GRS (full line). The dashed line represents the same sample of GRS clouds but with the assumption of non-sphericity of the clouds (This is performed by imposing that their radius in the direction of the outer Galaxy is $L/2$ and their radius in the direction of the inner Galaxy is given by $L/4+(gL)/4$) where $g$ is a Gaussian distributed random number with a mean of $0$ and a standard deviation of $1$). The triple dot-dashed lined corresponds to the sub-sample of gravitationally bound clouds with $\alpha_{vir} < 1$.}
\label{fig5}
\end{figure}

Using their measured gas surface densities, velocity dispersions, and estimated values of $A$, we now calculate the $S_{g}$ values of individual clouds using Eq.~\ref{eq8}. We adopt a value of $C=10^{3}$ ($\alpha_{A} \sim 3.45$), as in this work, we are concerned with the effects of shear on the scale of molecular clouds and the ability of shear to erase condensations before they are able to collapse into stars. This value of $C$ corresponds to the density contrast between the average molecular cloud density ($\sim 100$ cm$^{-3}$) and the density of $\sim 10^{5}$ cm$^{-3}$ at which molecular gas becomes gravitationally bound as pointed out by recent theoretical and observational findings (Dib et al. 2007; Parmentier 2011).  Fig.~\ref{fig5} (full line) displays the distribution of $S_{g}$ values for the entire sample of GRS clouds, while the dot-dashed line displays the distribution of $S_{g}$ values for clouds that are gravitationally bound (i.e., $\alpha_{vir} < 1$). In calculating the values of $A$ using Eq.~\ref{eq9}, we are making the assumption that clouds are roughly spherical as we consider their radii in the directions of the centre of the Galaxy and in the direction of the outer Galaxy to be equal. This is obviously an approximation as most Galactic molecular clouds are observed to have an aspect ratio in the range of $\sim [1.4-2.2]$ (e.g., Fig. 1 in Dib et al.  2009 based on the data of Heyer et al. 2001;  also Koda et al. 2006) and are likely to have a more complex 3D structure (Jones \& Basu 2002). We quantify the effects of dropping the sphericity assumption on the calculation of $A$ by: fixing the inner [or outer]  extent of the clouds at the Galactocentric position  (R-L/2) [or R+L/2] and assume that the clouds extent toward the outer [inner] Galaxy is given by $(R+(L/4)+(g L)/4)$ $[R-(L/4)-(g L)/4]$ where $g$ is a Gaussian distributed random number with a mean of $0$ and a standard deviation of $1$. One of these random realisations and its effect on the calculated values of $S_{g}$ is shown in Fig.~\ref{fig5} (dashed line histogram). While some variations can be noticed, the overall shape of the $S_{g}$ distribution is essentially unchanged. 

\begin{figure}
\plotone{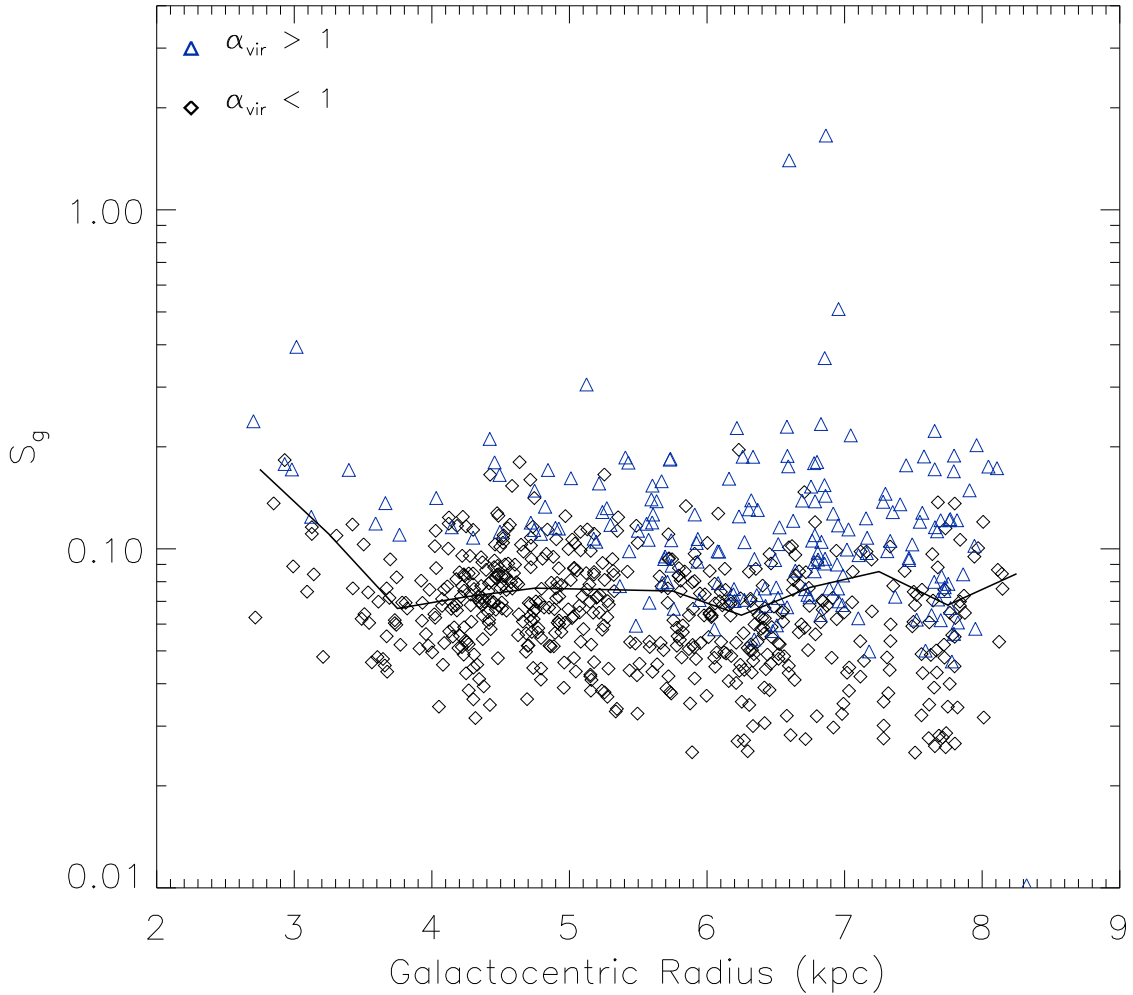}
\vspace{0cm}
\caption{The variation of the shear parameter $S_{g}$ for the selected sample of molecular clouds in the Galactic Ring Survey as a function of the Galactocentric Radius. Black diamonds are the sub-sample of gravitationally bound clouds (i.e., $\alpha_{vir} < 1$) and the blue triangles the sub-sample of unbound clouds (i.e., with $\alpha_{vir}  > 1$). The line displays the median value of $S_{g}$ for the entire sample of clouds in radial bins of 0.5 kpc.}
\label{fig6}
\end{figure}

Fig.~\ref{fig5} shows that almost all molecular clouds have $S_{g} < 1$ with the distribution peaking at $\sim 0.065$. This implies that Galactic shear is playing only a minor role in the overall global support of clouds against their own self-gravity and it also reflects the fact that almost all molecular clouds in the Galactocentric radius range covered by the GRS are prone to star formation. Fig.~\ref{fig6} displays the $S_{g}$ values of the clouds versus their Galactocentric radius. The gravitationally bound clouds ($\alpha_{vir} < 1$) are shown with the black diamonds and the unbound clouds ($\alpha_{vir} > 1$) with the blue triangles. This figure shows that there is no significant variation of $S_{g}$ with Galactocentric radius. 

\begin{figure}
\plotone{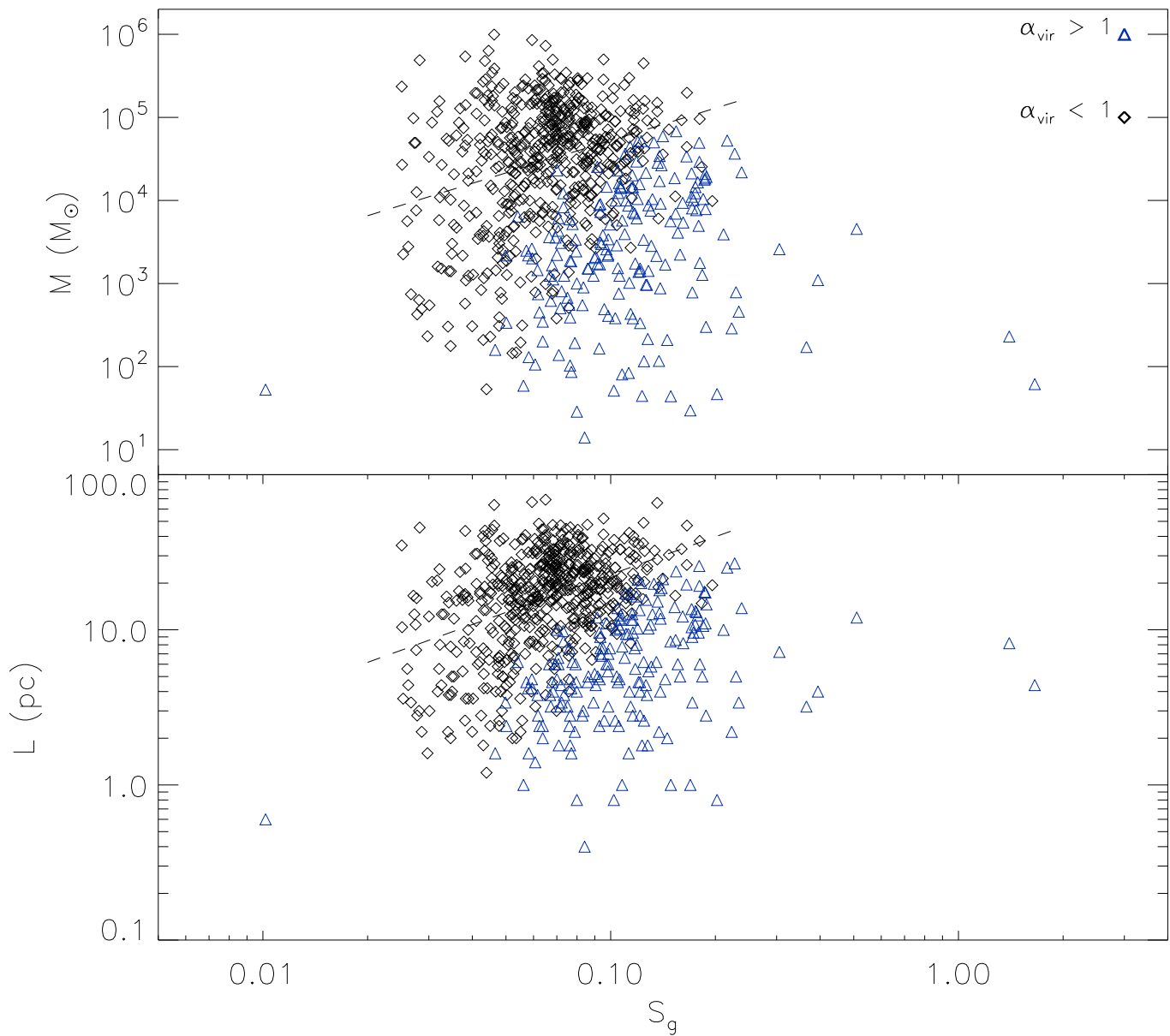}
\vspace{1.5cm}
\caption{The relationship between the shear parameter $S_{g}$ of the GRS clouds, and the masses (top), and sizes (bottom) of the clouds in the GRS. Empty diamonds designate the clouds with $\alpha_{vir} < 1$ and the empty triangles designate the clouds with $\alpha_{vir} > 1$. The dashed line corresponds to a linear fit to the $M-S_{g}$ data for the gravitationally bound clouds.}
\label{fig7}
\end{figure}

We now verify whether shear correlates with any of the clouds basic parameters. Fig.~\ref{fig7} displays the $S_{g}$ values of the gravitationally bound clouds (black diamonds) and the unbound ones (blue triangles) versus their masses (top) and sizes (bottom). Fig.~\ref{fig7} shows that for the bulk of the GRS molecular clouds, extending over 5 and 2.5 orders of magnitude in mass and physical size, there is no evidence of a strong correlation between their masses, sizes, and their shear levels. A calculation of the Spearman rank coefficient for the entire sample of clouds (bound+unbound) yield low correlation factors of $-0.064$ and $0.044$ for the $S_{g}-M$ and $S_{g}-L$ data, respectively. Weidner et al. (2010) and Escala (2011) argued that there is potentially a correlation between the mass of the most massive young cluster that can form in a galaxy (and therefore of its progenitor gaseous clump) and the global level of shear in the galaxy. This physical effect would still apply to an ensemble of clouds in a single galaxy. There might be some indication from Fig.~\ref{fig6} of the existence of an upper envelope in the $S_{g}-M$ relation in which the maximum value of $M$ decreases with increasing values of $S_{g}$. This trend is however not entirely conclusive due to the small number of clouds that are observed with $S_{g} > 0.3$. This may also reflect an observational bias since the density of these unbound clouds, which are likely to be dispersing, would fall below the threshold density necessary to excite the $^{13}$CO $J=1-0$ line. However, if one considers only the clouds that are the most likely progenitors of star clusters (i.e., with $\alpha_{vir} <1$, black diamonds), Fig.~\ref{fig7} does not suggest that $M$ decreases with increasing $S_{g}$. Instead, we observe the opposite trend which is one in which $S_{g}$ increases with increasing values of $M$ (and $L$) (the Spearman correlation coefficients are $0.22$ and $0.36$, respectively). A linear fit to the $S_{g}-M$ and $S_{g}-L$ relations for the sub-sample of bound clouds yields the following relations $M =10^{6.01 \pm 0.22}  S_{g}^{1.29 \pm 0.18}$ which is, given the weak dependence of $A$ on $M$ (i.e., Fig.~\ref{fig4}), and within the 1-$\sigma$ error bars, a reflection of the scaling relations of the clouds $M \propto S_{g}^{(\gamma \delta/\beta)} \sim S_{g}^{\sim 1.46}$. The increase of $M$ with $S_{g}$ for gravitationally bound clouds is an indication that their evolution and mass growth are unaffected by shear. 

\section{Correlation of Shear and Star Formation Efficiency Indicators}\label{results}

In this section, we explore the relationship between the shear that affects the clouds and indicators of their star formation efficiency. As pointed out in Moore et al. (2012), the ratio of the integrated luminosity to the mass of a molecular cloud, $L_{bol}/M$ is determined by its mean star-formation efficiency ($SFE$). The star formation efficiency is the star formation rate $SFR$ per unit gas mass multiplied by the timescale of star formation in the cloud ($L_{bol}/M) \propto SFE= \tau_{SF} (SFR/M)$. As discussed in Moore et al. (2012), a high value of $L_{bol}/M$ can be produced by either a high $SFR$ per unit gas mass ($SFR/M$) or by a long timescale of star formation ($\tau_{SF}$). In the case of Galactic GMC, the timescale sampled by the data is limited to the lifetimes of the evolutionary stages traced by RMS. These lifetimes are those of the MYSOs and compact  \ion{H}{ii}-region stages, the durations of which have both been determined to be $< 10^{6}$ yr by Mottram et al. (2011). These timescales are shorter than the typical crossing time of the GRS clouds of $\sim 3.2 \times 10^{6}$ yr that is defined as being the ratio of the median size to median velocity dispersion of the clouds, and thus, they imply that the RMS data do not trace multiple generations of star formation but provides a snapshot of the {\it current} $SFE$. The use of $(L_{bol}/M)$ as a proxy of the $SFE$ is valid as long as the stellar initial mass function (IMF) and as a consequence the luminosity function are invariant. Variations to the IMF such as those discussed in Dib et al. (2010b; Dib 2012) in regions of massive star formation may invalidate the underlying linearity assumption of the $(L_{bol}/M)-SFE$ relationship. Fig.~\ref{fig8} displays the relationship between the shear parameter $S_{g}$ and $(L_{bol}/M)$ for a subsample of 125 GRS clouds for which MYSOs luminosities are available from the Red MSX survey of Urquhart et al. (2011). A calculation of the Spearman correlation coefficient of the $(L_{bol}/M)-S_{g}$ yields a value of $0.09$ which is an indication of a nonsignificant correlation between the two quantities.

Using the GRS published data, it is possible to measure another indicator of the clouds efficiency in converting their gas reservoirs into stars. Rathborne et al. (2010) established the list of clumps in each GRS cloud, but their masses were not estimated. We use the formalism presented in Simon et al. (2001) along with the available information on the clump population and measure the masses of the individual clumps in the GRS clouds. The mass of a clumps is given by:

\begin{equation}
M_{clu}=3.05 \times 10^{-25} N(^{13}{\rm CO}) \theta_{x} \theta_{y} D^{2}, 
\label{eq10}
\end{equation}

where $D$ is the distance to the cloud harbouring the clump in parsecs, $\theta_{x}$ and $\theta_{y}$ are the sizes of the principal half axis in arcseconds, and N$(^{13}$CO) is the total $^{13}{\rm CO}$ column density (in cm$^{-2}$). The quantity N($^{13}$CO) can be estimated using (Simon et al. 2001):

\begin{equation}
N(^{13}{\rm CO})= 8.75 \times 10^{14} T_{mb} \Delta v,
\label{eq11}
\end{equation}

where $T_{mb}$ is the brightness temperature of the clumps (in K) and $\Delta v$ their FWHM line widths (in km s$^{-1}$) in the optically thin, thermalized limit. Using the values of $\theta_{x}$, $\theta_{y}$, $T_{mb}$, and $\Delta v$, derived by Rathborne et al. (2010), we calculate the masses of all the clumps ($\sim 6100$) that are identified in 518 molecular clouds in the GRS. For each GRS cloud that possesses one or more associated clumps, we measure the quantity $\Sigma_{i} M_{clu,i}/M$, which is the ratio of the total mass of the clumps in a given cloud to the cloud mass. As various clumps in a cloud are likely to have different ages and be in different evolutionary/contraction stage, the quantity $\Sigma_{i} M_{clu,i}/M$ can be viewed as a lower limit to the clump formation efficiency  $CFE_{clu}$ (i.e., in the case no further clump formation occurs before the cloud is disrupted). Fig.~\ref{fig9} displays the dependence of $CFE_{clu}$ as a function of the shear parameter. Here again, no significant correlation is observed between the $CFE_{clu}$ and $S_{g}$ (Spearman correlation coefficient of -0.08). We also explore whether lower levels of shear are associated with higher levels of fragmentation in the clouds as would be expected if shear was playing a significant role in the dynamics of the clouds. Fig.~\ref{fig10} displays the shear parameter versus the number of clumps found in the clouds, $N_{clu}$ (top, black diamonds represents bound clouds and blue triangles represent unbound clouds) and the number of clumps per unit mass, $N_{clu}/M$ (bottom). The Spearman correlation coefficients between $S_{g}$ and $N_{clu}$ and $S_{g}$ and $N_{clu}/M$ are, for the entire sample, $\sim -0.15$ and $-0.03$, respectively, while for the population of gravitationally bound clouds the coefficients are $-0.01$ and $-0.24$. These values are indicative of weak to very weak correlations between the shear parameter of the clouds and their levels of fragmentation.  

Figs.~\ref{fig8}, \ref{fig9}, and \ref{fig10} show that there is essentially no significant correlations between the shear parameters of individual clouds and several indicators of their star formation activity. However, it is important to consider that any sample of Galactic molecular clouds such as the GRS will contain clouds, in a given mass range, that are in various evolutionary stages. Hence, the true correlation between the SFR and SFE of a cloud and the level of shear it is subjected to may only appear when an integration is made over the entire cloud lifetime. Alternatively, a similar putative correlation will be present between shear and star formation rates by averaging over the properties  of large samples. As we are attempting to test the numerical results of Weidner et al. (2010), the expected trend if the effect of shear was important are lower $S_{g}$ values for the higher SFRs that are observed towards the centre of the Galaxy. Thus, we would expect a correlation between $S_{g}^{-1}$ and the SFR and SFE. Fig.~\ref{fig11} (top) displays the dependence of the median value of $S_{g}^{-1}$ for the GRS clouds as a function of the Galactocentric radius in radial bins of 0.5 kpc. The filled symbols in Fig.~\ref{fig11} (top) show the Galactic radial profile of the SFR from G\"{u}sten \& Mezger 1983 (filled triangles), Lyne et al. 1985 (filled squares) and Guibert et al. 1978 (filled squares) and the radial profile of the MYSOs luminosity surface density, $\Sigma_{L_{bol}}$, in the Galactic sector of the GRS and which is a proxy of the SFR (from Moore et al. 2012). All quantities are normalised to their respective values at the Galactocentric radius of the Sun. Fig.~\ref{fig11} (bottom) compares the same normalized radial median values of $S_{g}^{-1}$  to the ratio of the total RMS source luminosity to the total mass in GRS clouds ($L_{bol}/M$) as a function of Galactocentric radius in radial bins of 0.5 kpc. A two-sided Kolmogorov-Smirnov (K-S) test indicates that the $S_{g}^{-1}-SFR$ and the $S_{g}^{-1}-\Sigma_{L_{bol}}$ distributions are likely to be drawn from the same distribution with probabilities of only $1.7 \times 10^{-4} \%$ and $5.3\times 10^{-5} \%$. A similar K-S test to the $S_{g}^{-1}-(L_{bol,tot}/M_{tot})$ distributions indicates that they have a probability of only $0.14 \%$ to be drawn from the same distribution. 

\section{Discussion and Conclusions}\label{conclusions}

Recent numerical simulations by Weidner et al. (2010) of star cluster formation in a $10^{6}$ M$_{\odot}$ cloud suggests that the star formation efficiency (SFE) depends on the initial rotational velocities of the cloud. In this work, we used data from the Galactic Ring Survey to test the importance of Galactic shear in regulating the SFE on the scale of individual molecular clouds ($\sim 730$ clouds with sizes in the range of 0.2-35 pc ). We calculate the shear parameter, $S_{g}$, which is the ratio between a critical surface density for perturbations in molecular clouds to grow by a factor $10^{3}$ (thus to reach densities of $\sim 10^{5}$ cm$^{-3}$ and become gravitationally bound) in the presence of shear to the actual surface density of the clouds. We find that the distribution of $S_{g}$ is peaked around a value that is smaller than unity ($\sim 0.065$). Since molecular clouds in the GRS sample are likely to be in various evolutionary stages in terms of their gravitational boundedness, this suggests that Galactic shear plays only a minor role, at any given time, in providing a global support against gravity in these clouds. It is also an indication that molecular clouds can only survive in low shear environments. We also find that there is no dependence of the $S_{g}$ values on Galactocentric radius. We also find that for gravitationally bound clouds, which are the most likely to be the progenitors of star clusters, there is a tendency for the $S_{g}$ values to increase with increasing mass. This casts some doubts on the idea that the maximum mass of a cluster is limited by the local level of shear.

\begin{figure}
\plotone{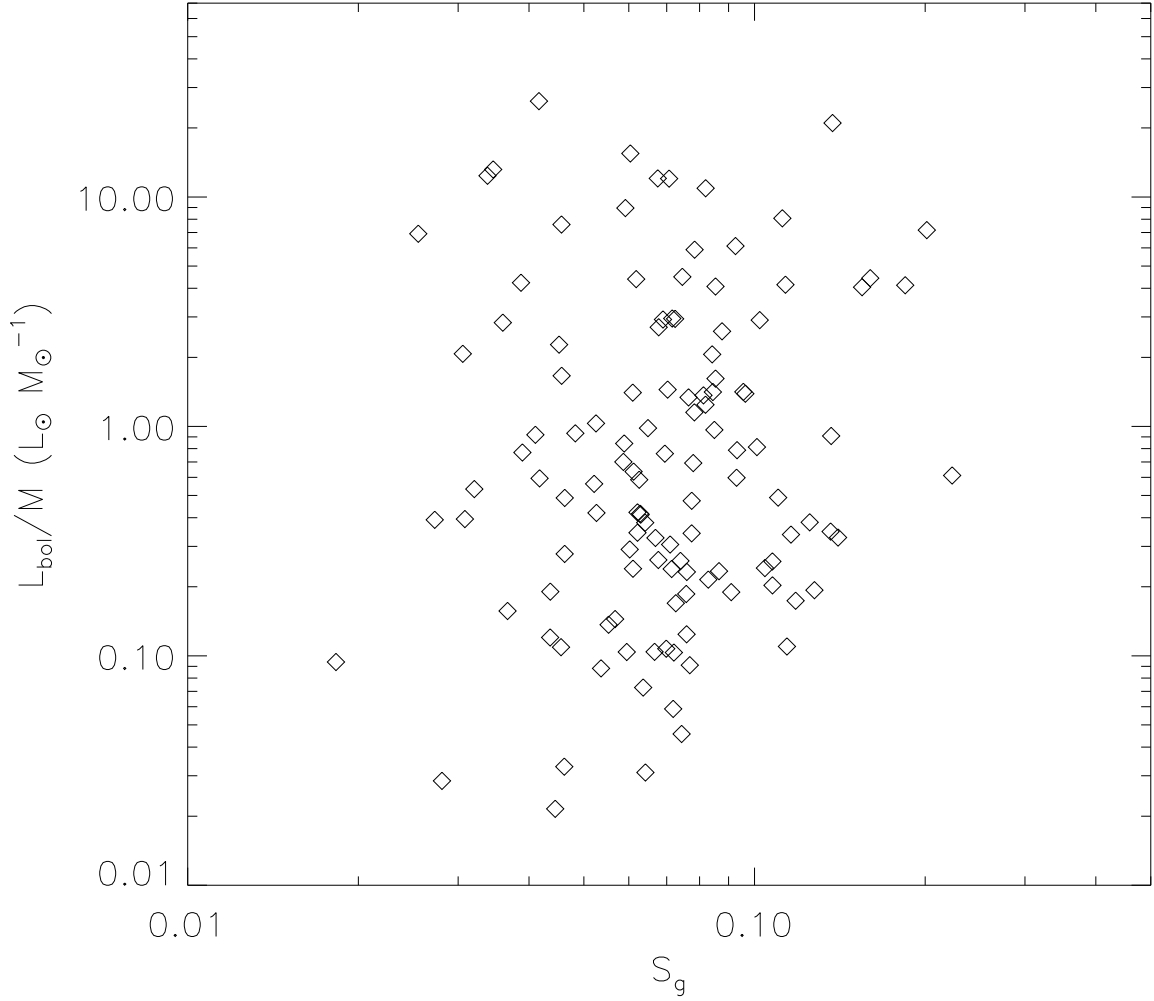}
\caption{The relationship between the shear parameter $S_{g}$ of the GRS clouds and the quantity $L_{bol}/M$ which is a proxy for the star formation efficiency in the clouds. The bolometric $L_{bol}$ luminosity is measured for massive YSOs in the Red MSX survey (Urquhart et al. 2011).}
\label{fig8}
\end{figure}

\begin{figure}
\plotone{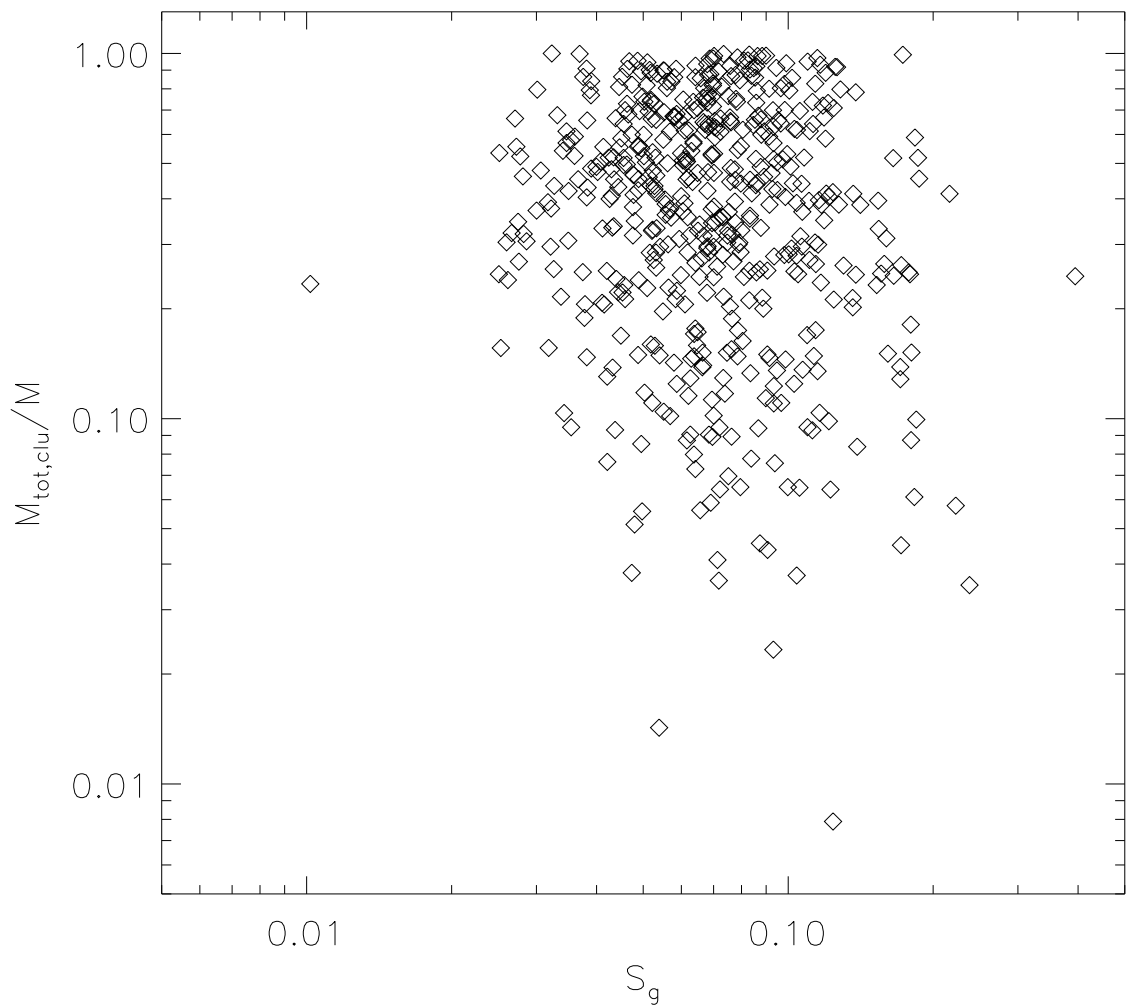}
\caption{The relationship between the shear parameter $S_{g}$ of the GRS clouds and the quantity  $M_{tot,clu}/M$ which is a proxy for the clump formation efficiency in the clouds. The quantity $M_{tot,clu=} \Sigma_{i} (M_{i,clu})$ is the sum of the masses of the clumps present in cloud $i$.}
\label{fig9}
\end{figure}

\begin{figure}
\plotone{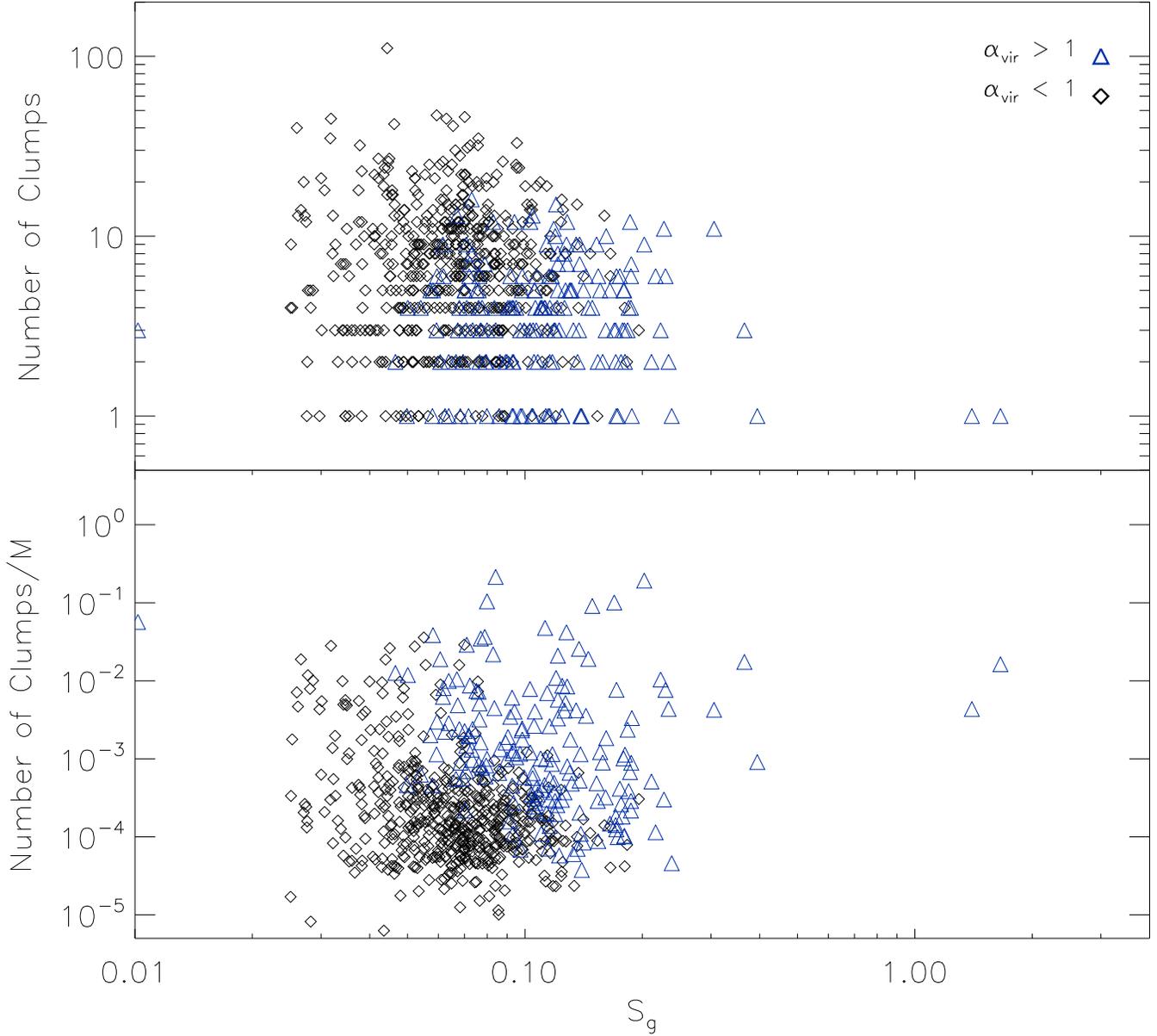}
\vspace{1.5cm}
\caption{Correlation between the $S_{g}$ parameter of the GRS clouds and the number of clumps they harbour (top) and the number of clumps per unit mass (bottom) for the gravitationally bound clouds (black diamonds) and unbound clouds (blue trianges).}
\label{fig10}
\end{figure}

\begin{figure}
\plotone{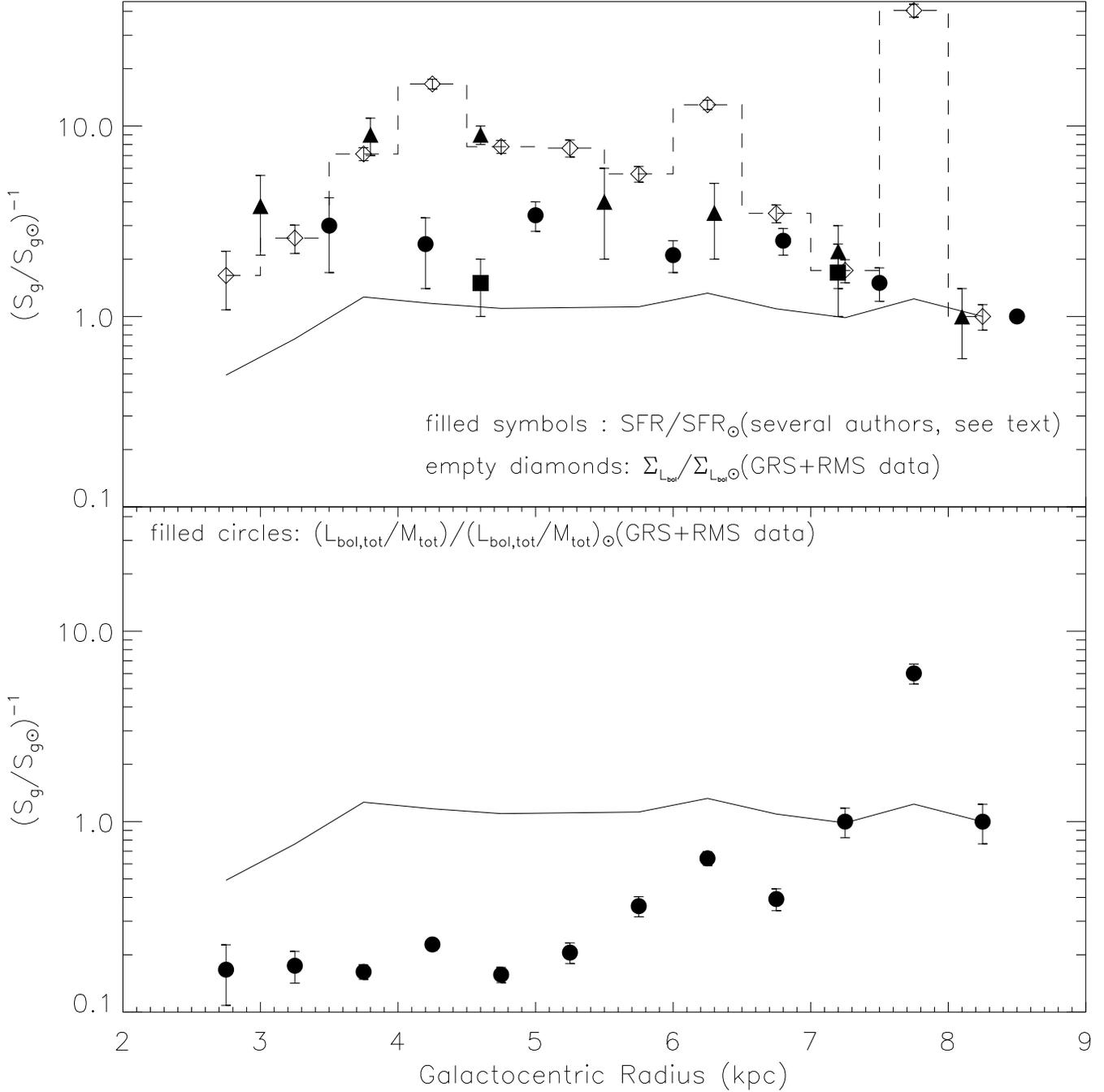}
\vspace{1.5cm}
\caption{The dependence of $S_{g}^{-1}$ as a function of the Galactocentric radius is compared, in radial bins of $0.5$ kpc (top panel) to that of the SFR using the data from Guibert et al. (1978, filled squares), G\"{u}sten \& Mezger (1982, filled triangles), and Lyne et al. (1985, filled circles) and to the surface density of the luminosity of the MYSOs, $\Sigma_{L_{obl}}$ which are found in the GRS sector of the Galaxy (empty diamonds). The bottom panel compares the radial depedence of $S_{g}^{-1}$ to the ratio of the total MYSOs luminosity to the total mass of the clouds in the GRS. All quantities are normalized to their respective value at the Galactoncentric radius of the Sun.}
\label{fig11}
\end{figure}

In order to test the effects of shear on the SFE on the scale of individual clouds, we search for correlations between $S_{g}$ and the quantity $(L_{bol}/M)$ where $M$ is the mass of the cloud and $L_{bol}$ is the bolometric luminosity of the massive young stellar objects (MYSOs) measured for a sub-sample of the GRS clouds observed in the Red MSX survey (Urquhart et al. 2011). Under the assumption that the luminosity function is invariant in the Galaxy, the ratio $(L_{bol}/M)$ is a suitable proxy of the SFE (e.g., Moore et al. 2012). No significant correlation is found between the clouds $S_{g}$ values and their SFE (i.e., Fig.~\ref{fig8}) which implies that shear is unlikely to be a dominant mechanism which determines the SFE of clouds. We also compare the $S_{g}$ values of the clouds to the fraction of their mass that is found in the smaller and denser clumps they harbour and which is a proxy of the clouds clump formation efficiency (CFE), and between the $S_{g}$ values and the number of clumps they harbour per unit mass (Number of clumps/$M$). The insignificant correlations that are found between the $S_{g}$ values and the CFE (Fig.~\ref{fig9}) and between the $S_{g}$ values of the clouds and their level of fragmentation (Fig.~\ref{fig10}) also suggest that shear is playing a negligible role in determining the rate and efficiencies of star formation in molecular clouds. Finally, we compare the radial distributions of the mean values of $S_{g}$ (in effect the distribution of $S_{g}^{-1}$ as stronger shear, if dominant, is expected to reduce the SFE) to that of the radial distributions of the SFR and SFE using both data from the literature and from the Red MSX survey. We find that the $S_{g}^{-1}$ distribution and the radial distributions of the the SFR and SFE have very low probabilities of being drawn from the same underlying distribution.

In contrast, observations in the \ion{H}{i} 21 cm line in a number of external galaxies on physical scales of $\sim 300-400$ pc by Hunter et al. (1998) and Elson et al. (2012) indicate the existence of positive correlations between the radial distribution of $S_{g}$ values measured on these scales and the radial extent of the stellar disks in those galaxies. These results suggest that shear regulates the formation of molecular clouds  and thus helps determine {\it where} can molecular clouds form in a galaxy. Current observations do not allow the measurement of the SFE in external galaxies on scales of $300-400$ pc. However,  observations by Seigar (2005) do not suggest the existence of significant correlations between the global SFR values in galaxies and their global levels of shear. While the results of Hunter et al. (1998) and Elson et al. (2012) that shear determines the {\it spatial extent} of the molecular star forming region in galactic disks, our results suggest that shear does not influence the {\it efficiency} at which stars form in individual molecular clouds.

Our interpretation is that the results found by Weidner et al. (2010), i.e., higher SFEs in clouds with lower initial levels of rotation stem from the neglect of a number of physical processes. It is important to stress that while shear may participate in determining the spatial extent of where molecular clouds (and as consequence stars) can form in galactic disks (e.g., Hunter et al. 1998; Elson et al. 2012), it is not the only culprit. Blitz \& Rosolowsky (2006) have demonstrated that there is a relatively tight correlation between the ratio of molecular-to-neutral hydrogen and the local pressure in the disks $P_{ext}$. As discussed by Blitz \& Rosolowsky (2006), $P_{ext}$ is dominated by the density of stars in most galactic disks. This seems to suggest that stars already present in the disk influence the formation of newer generations of molecular clouds and as consequence of newer generations of stars. The role of existing stars in regulating the SFE in galaxies has also been explored recently by Shi et al. (2011) who found tighter correlations between the surface density of star formation and the surface densities of both gas and stars than with the surface density of gas alone using observations on sub-kpc scales in 12 nearby galaxies. The models of Weidner et al. (2010) did not consider the effects of magnetic fields and stellar feedback. Magnetic fields are known to strongly regulate the rate at which molecular clouds convert a fraction of their mass into dense clumps and cores per unit time. Numerical simulations of magnetised molecular clumps and clouds by several groups (e.g., V\'{a}zquez-Semadeni et al. 2005; Price \& Bate 2008; Dib et al. 2010a; Li et al. 2010) have repeatedly shown, using different numerical techniques, that stronger magnetic fields in terms of magnetic criticality reduce significantly the efficiency at which clumps/clouds convert their gas reservoir into dense cores. Price \& Bate (2008) observed a $75 \%$ reduction in the star formation efficiency measured after $1.5$ the free-fall time of the clump when going from a clump with a weak magnetic field (i.e., mass-to-magnetic flux ratio, $\mu=20$) to one with stronger magnetic field ($\mu=3$). Dib et al. (2010a) derived the core formation efficiency per unit free-fall time of the molecular cloud $CFE_{ff}$ in their simulations of a magnetised, self-gravitating clouds with decaying turbulence and found that the $CFE_{ff}$ varies from $6 \%$ for a cloud with $\mu=3$ to $33 \%$ for $\mu=9$. Supersonic turbulence, which is ubiquitously observed in molecular clouds also affects the rates and efficiencies with which clouds convert their gas into dense cores and stars. Using numerical simulations of magnetised and turbulent clouds, Padoan \& Nordlund (2011) showed that the star formation rate per unit free-fall time of the cloud, $SFR_{ff}$, depends on the Mach number and virial parameter of the cloud. However, Murray (2011) recently estimated the values of the star efficiency per free-fall time, $SFE_{ff}$, in a number of massive star forming Galactic GMCs and found that the $SFE_{ff}$ in a subsample of the most actively star forming GMCs is much higher than the mean $SFE_{ff}$ of the entire sample. As the dynamical condition in these GMCs (i.e., rms Mach number and virial parameter) are not substantially different, Murray concluded that turbulence cannot be the dominant process that regulates the $SFE_{ff}$ in massive Galactic GMCs. Once the first generation of stars form in the clump/cloud, feedback from stars and particularly from massive stars is expected to influence the ability of the clump/cloud to form further generations of stars. Dib et al. (2011) recently presented a model which describes star formation in protocluster clumps in which the star formation process is stopped by the expulsion of gas from the clump by the winds of OB stars. They showed that variations of the $CFE_{ff}$ by a factor of 3 in the cloud (that can be attributed to variations in the level of turbulence and magnetic field strength) result in variations of the final SFE by only a factor of $0.6$ by the time gas is expelled from the protocluster clump. This is due to the highly non-linear dependence of stellar wind feedback on the stellar mass $\propto M_{*}^{4}$. Obviously, more observational and theoretical efforts are needed in order to measure the relative importance of the different mechanisms that regulate the formation of dense gravitationally bound cores in clumps/clouds, and as a consequence the efficiency of star formation, at various epochs of the GMCs lifetimes.  

\acknowledgements

We are very grateful to the referee for a thoughtful report which helped us improve both the content of the paper and the presentation of the results. We also thank Samuel Boissier and Ruixiang Chang for useful discussions. SD acknowledges support from the Science and Technology Facilities Council (STFC) grant ST/H00307X/1 and a Santander Mobility Award. TJTM is supported in part by STFC grant ST/001847/1. This publication makes use of molecular line data from the Boston University-FCRAO Galactic Ring Survey (GRS). The GRS is a joint project of Boston University and Five College Radio Astronomy Observatory, funded by the National Science Foundation under grants AST-9800334, AST-0098562, \& AST-0100793. This paper made use of information from the Red MSX Source survey database at www.ast.leeds.ac.uk/RMS which was constructed with support from the STFC. 
  
{}

\newpage

\newpage

\newpage

\newpage

\newpage

\newpage

\begin{thebibliography}{}

\bibitem[Andre (2009)] {andre09} Andr\'{e}, P., Basu, S., \& Inutsuka, S. \ 2009, Structure Formation in Astrophysics, Ed G. Chabrier, Cambridge Univ. Press, 254 
\bibitem[Ballesteros (2006)] {ballesteros06} Ballesteros-Paredes, J. \ 2006, \mnras, 372, 443 
\bibitem[Blitz (1993)] {blitz93} Blitz, L. \ 1993, in Protostars and Planets III, eds. E. H. Levy \& J. I. Lunine (Tuscon, AZ: Univ. Arizona Press), 125 
\bibitem[Blitz (2006)] {blitz06} Blitz, L., \& Rosolowsky, E. \ 2006,  \apj, 650, 933 
\bibitem[Clemens (1985)] {clemens85} Clemens, D. P. \ 1985, \apj, 295, 422
\bibitem[Csengeri (2011)] {csengeri11} Csengeri, T., Bontemps, S., Schneider, N., Motte, F., \& Dib, S. \ 2011, \aap, 527, 135 
\bibitem[Dib (2005)] {dib05a} Dib, S. \ 2005, PhD Thesis, Combined Faculties for the Natural Sciences and for Mathematics of the University of Heidelberg, Germany
\bibitem[Dib (2005)] {dib05b} Dib, S., \& Burkert, A. \ 2005, \apj, 630, 238
\bibitem[Dib (2006)] {dib06} Dib, S., Bell, E., \& Burkert, A. \ 2006, \apj, 638, 797
\bibitem[Dib (2007a)] {dib07a} Dib, S., Kim, J., V\'{a}zquez-Semadeni, E., Burkert, A., \& Shadmehri, M. \ 2007, \apj, 661, 262
\bibitem[Dib (2007b)] {dib07b} Dib, S., Kim, J. \ 2007, in Haverkorn M., Goss, M. eds, ASP Conf. Ser. Vol. 365, Small Ionized and Neutral Structures in the Diffuse Interstellar Medium. Astron. Soc. Pac. San Francisco, p. 166
\bibitem[Dib (2008)] {dib08} Dib, S., Brandenburg, A., Kim, J., Gopinathan, M.,\& Andr\'{e}, P. \ 2008, \apj, 678, L105
\bibitem[Dib (2009)] {dib09} Dib, S., Walcher, C. J., Heyer, M., Audit, E., \& Loinard, L. \ 2009, \mnras, 398, 1201
\bibitem[Dib (2010a)] {dib10a} Dib, S., Hennebelle, P., Pineda, J. E., Csengeri, T., Bontemps, S., Audit, E., \& Goodman, A. A. \ 2010a, \apj, 723, 425
\bibitem[Dib (2010b)] {dib10b} Dib, S., Shadmehri, M., Padoan, P., Maheswar, G., Ojha, D. K., Khajenabi, F. 2010b, \mnras, 405, 401  
\bibitem[Dib (2011a] {dib11a} Dib, S., Piau, L., Mohanty, S., \& Braine, J. \ 2011, \mnras, 415, 3439
\bibitem[Dib (2011b)] {dib11b} Dib, S. \ 2011, \apjl, 737, L20 
\bibitem[Dib (2012)] {dib12} Dib, S. \ 2012 in The Labyrinth of Star Formation, eds. D. Stamatellos, S. Goodwin, \& D. Ward-Thompson. Springer, in press, (arXiv:1207.4982)
\bibitem[Dobbs (2008)] {dobbs08} Dobbs, C. L. \ 2008, \mnras, 391, 844
\bibitem[Elmegreen (1987)] {elmegreen87} Elmegreen, B. G. \ 1987, \apj, 312, 626
\bibitem[Elmegreen (1993)] {elmegreen93} Elmegreen, B. G. \ 1993, in Star Formation, Galaxies, and the Interstellar Medium, ed. J. Franco et al. (Cambridge: Cambridge Univ. Press), 337 
\bibitem[Elmegreen (1995)] {elmegreen95} Elmegreen, B. G. in Molecular Clouds Star Formation, eds. C. Yuan \& Y.-H. You p.149, Singapore, World Sci 
\bibitem[Elson (2012)] {elson12} Elson, E. C., de Block, W. J. G., \& Kraan-Korteweg, R. C. \ 2012, \apj, 143, 1
\bibitem[Escala (2011)] {escala11} Escala, A. \ 2011, \apj, 735, 56
\bibitem[Goodwin (2004)] {goodwin04} Goodwin, S. P., Whitworth, A. P., \& Ward-Thompson, D. \ 2004, \aap, 423, 169
\bibitem[Guibert (1978)] {guibert78} Guibert, J., Lequeux, J., \& Viallefond, F. \ 1978, \aap, 68, 1
\bibitem[Gusten (1983)] {gusten82} G\"{u}sten, R., \& Mezger, M. \ 1982, Vistas Astron., 26, 159
\bibitem[Heitsch (2009)] {heitsch09} Heitsch, F., Stone, J. M., Hartmann, L. \ 2009, \apj, 695, 248 
\bibitem[Heyer (2001)] {heyer01} Heyer, M. H., Carpenter, J. M. \& Snell, R. L. \ 2001, \apj, 551, 852
\bibitem[Heyer (2004))] {heyer04} Heyer, M. H., \& Brunt, C. M. \ 2004, \apjl, 615, L45
\bibitem[Hocuk (2011)] {hocuk11} Hocuk, S., \& Spaans, M. \ 2011, \aap, 536, 41
\bibitem[Hunter (1998)] {hunter98} Hunter, D. A., Elmegreen, B. G., \& Baker, A. L. \ 1998, \apj, 493, 595
\bibitem[Imara (2011)] {imara11a} Imara, N., \& Blitz, L. \ 2011, \apj, 732, 78
\bibitem[Imara (2011)] {imara11b} Imara, N., Bigiel, F., \& Blitz, L. \ 2011, \apj, 732, 79 
\bibitem[Jackson (2006)] {jackson06} Jackson, J. M., Rathborne, J. M., Shah, R. Y., et al. \ 2006, \apjs, 163, 145
\bibitem[Jones (2002)] {jones02} Jones, C. E., \& Basu, S. \ 2002, \apj, 569, 280 
\bibitem[Kerton (2003)] {kerton03} Kerton, C. R., Brunt, C. M., Jones, C. E., \& Basu, S. \ 2003, \aap, 411, 149  
\bibitem[Khesali (2007)] {khesali07} Khesali, A.-R., \& Bagherian, M.-A. \ 2007, Bull. Astron. Soc. India, 35, 1
\bibitem[Kim (2011)] {kim11} Kim, C.-G., Kim, W.-T. \& Ostriker, E. C. \ 2011, \apj, 743, 25
\bibitem[Klessen (2000)] {klessen00} Klessen, R. S., Heitsch, F., \& Mac Low, M.-M. \ 2000, \apj, 535, 887 
\bibitem[Koda (2006)] {koda06} Koda, J., Sawada, T., Hasegawa, T., \& Scoville, N. \ 2006, \apj, 638, 191
\bibitem[Koda (2009)] {koda09} Koda, J., Scoville, N., Sawada, T. et al. \ 2009, \apj, 700, L132
\bibitem[Kritsuk (2011)] {kritsuk11} Kritsuk, A. G., Norman, M. L., \& Wagner, R. \ 2011, \apjl, 727, L20 
\bibitem[Li (2006)] {li06} Li, Y., Mac Low, M.-M., \& Klessen, R. S. \ 2006, \apj, 639, 879 
\bibitem[Li (2010)] {li10} Li, Z.-Y., Wang, P., Abel, T., \& Nakamura, F. \ 2010, \apj, 720, L26  
\bibitem[Lyne (1985)] {lyne85} Lyne, A., Manchester, R., \& Taylor, J. \ 1985, \mnras, 213, 613
\bibitem[Madore (1977)] {madore77} Madore, B. F. \ 1977, \mnras, 178, 1 
\bibitem[Moore (2012)] {moore12} Moore, T. J. T., Urquhart, J. S., Morgan, L. K., \& Thompson, M. A. \ 2012, \mnras, accepted, in press, (arXiv:1204.1578) 
\bibitem[Mottram (2011)] {mottram11} Mottram, J. C., Hoare, M. G., Urquhart, J. S., Lumsden, S. L., Oudmaijer, R. D., Robitaille, T. P., Moore, T. J. T., Davies, B., Stead, J. \ 2011, \aap, 525, 149
\bibitem[Murray (2010)] {murray10} Murray, N., Quataret, E., \& Thompson, T. A. \ 2010, \apj, 709, 191
\bibitem[Murray (2011)] {murray11} Murray, N. \ 2011, \apj, 729, 133
\bibitem[Offner (2008)] {offner08} Offner, S. R., Klein, R. I., \& McKee, C. F. \ 2008, \apj, 686, 1174
\bibitem[Padoan (2011)] {padoan11} Padoan, P., \& Nordlund, \AA. \ 2011, \apj, 730, 40  
\bibitem[Parmentier (2011)] {parmentier11} Parmentier, G. \ 2011, \mnras, 413, 1899
\bibitem[Price (2008)] {price08} Price, D. J., \& Bate, M. R. \ 2008, \mnras, 385, 1820
\bibitem[Rathborne (2009)] {rathborne09} Rathborne, J. M., Johnson, A. M., Jackson, J. M., Shah, R. Y., \& Simon, R. \ 2009, \apjs, 182, 131
\bibitem[Robitaille (2006)] {robitaille06} Robitaille, T. P., Whitney, B. A., Indebetouw, R., Wood, K., Denzmore, P. \ 2006, \apjs, 167, 256 
\bibitem[Roman-Duval (2009)] {romanduval09} Roman-Duval, J., Jackson, J. M., Heyer, M., Johnson, A., Rathborne, J., Shah, R., \& Simon, R. \ 2009, \apj, 699, 1153
\bibitem[Roman-Duval (2010)] {romanduval10} Roman-Duval, J., Jackson, J. M., Heyer, M., Rathborne, J., \& Simon, R. \ 2010, \apj, 723, 492 
\bibitem[Romeo (2010)] {rome010} Romeo, A. B., Burkert, A., \& Agertz, O. \ 2010, \mnras, 407, 1223 
\bibitem[Rosolowsky (2007)] {rosolowsky07} Rosolowsky, E. \ 2007, \apj, 654, 240 
\bibitem[Schneider (2011)] {schneider11} Schneider, N., Bontemps, S., Simon, R., Ossenkopf, V., Federrath, C., Klessen, R. S., Motte, F., Andr\'{e}, Ph., Stutzki, J., \& Brunt, C. \ 2011, \aap, 529, 1 
\bibitem[Seigar (2005)] {seigar05} Seigar, M. S. \ 2005, \mnras, 361, L20 
\bibitem[Shadmehri (2010)] {shadmehri10} Shadmehri, M., Nejad-Asghar, M., \& Khesali, A. \ 2010, \apss, 326, 83 
\bibitem[Shadmehri (2011)] {shadmehri11} Shadmehri, M., \& Khajenabi, F. \ 2012, \mnras, 421, 841
\bibitem[Shetty (2012)] {shetty10} Shetty, R., Collins, D. C., Kauffmann, J., Goodman, A. A., Rosolowsky, E. W., \& Norman, M. L. \ 2010, \apj, 712, 1049
\bibitem[Shi (2011)] {shi11} Shi, Y., Helou, G., Lin, Y., Armus, L., Wu, Y., Papovivh, C., Stierwalt, S. \ 2011, \apj, 733, 87
\bibitem[Silk (1997)] {silk97} Silk, J. \ 1997, \apj, 481, 703
\bibitem[Simon (2001)] {simon01} Simon, R., Jackson, J. M., Clemens, D. P., Bania, T. M., \& Heyer, M. H. \ 2001, \apj, 551, 747
\bibitem[Slyz (2005)] {slyz05} Slyz, A. D., Devriendt, J. E. G., Bryan, G., \& Silk, J. \ 2005, \mnras, 356, 737
\bibitem[Solomon (1987)] {solomon87} Solomon, P. M., Rivolo, A. R., Barrett, J., \& Yahil, A. \ 1987, \apj, 319, 730
\bibitem[Tan (2000)] {tan00} Tan, J. C. \ 2000, \apj, 536, 173 
\bibitem[Toomre (1964)] {toomre64} Toomre, A. \ 1964, \apj, 139, 1217 
\bibitem[Urquhart (2011)] {urquhart11} Urquhart, J. S., Moore, T. J. T., Hoare, M., Lumsden, S. L., Oudmaijer, R. D., Rathborne, J. M., Mottram, J. C., Davies, B., Stead, J. J. \ 2011, \mnras, 410, 1237
\bibitem[Vazquez (2005)] {vazquez05} V\'{a}zquez-Semadeni, E., Kim, J., \& Ballesteros-Paredes, J. \ 2005, \apj, 630, L49
\bibitem[Wada (2000)] {wada00} Wada, K., Spaans, M., \& Kim, S. \ 2000, \apj, 540, 797
\bibitem[Watson (2012)] {watson12} Watson, L. C., Martini, P., Lisenfeld, U., Wong, M.-H., B\"{o}ker, T., Schinnerer, E. \ 2012, \apj, 751, 123
\bibitem[Weidner (2010)] {weidner10} Weidner, C., Bonnell, I. A., \& Zinnecker, H., \ 2010, \apj, 724, 1503  
\bibitem[Williams (1994)] {williams94} Williams, J. P., de Geus, E. J., \& Blitz, L. \ 1994, \apj, 428, 693 
\bibitem[Wong (2002)] {wong02} Wong, T. \& Blitz, L. \ 2002, \apj, 569, 157 


\end{thebibliography}
\end{document}